\documentclass{article}
\usepackage{arxiv}

\usepackage[utf8]{inputenc} 
\usepackage[T1]{fontenc}    

\usepackage{hyperref}       
\usepackage{url}            
\usepackage{booktabs}       
\usepackage{amsfonts}       
\usepackage{nicefrac}       
\usepackage{microtype}      
\usepackage{graphicx}
\usepackage{amsmath}
\usepackage{multirow}
\usepackage{amsthm}

\theoremstyle{remark}

\title{Time-varying Network Driver Estimation (TNDE) quantifies stage-specific regulatory effects from single-cell snapshots}

\author{
 Jiaxin Li$^1$, Shanjun Mao$^{1\dagger}$ \\
  $^1$Department of Statistics and Data Science, Hunan University\\
  $\dagger$: To whom correspondence should be addressed.
}

\begin{document}

\maketitle
\begin{abstract}
Identifying key driver genes governing biological processes such as development and disease progression remains a challenge. While existing methods can reconstruct cellular trajectories or infer static gene regulatory networks (GRNs), they often fail to quantify time-resolved regulatory effects within specific temporal windows. Here, we present Time-varying Network Driver Estimation (TNDE), a computational framework quantifying dynamic gene driver effects from single-cell snapshot data under a linear Markov assumption. TNDE leverages a shared graph attention encoder to preserve the local topological structure of the data. Furthermore, by incorporating partial optimal transport, TNDE accounts for unmatched cells arising from proliferation or apoptosis, thereby enabling trajectory alignment in non-equilibrium processes. Benchmarking on simulated datasets demonstrates that TNDE outperforms existing baseline methods across diverse complex regulatory scenarios. Applied to mouse erythropoiesis data, TNDE identifies stage-specific driver genes, the functional relevance of which is corroborated by biological validation. TNDE offers an effective quantitative tool for dissecting dynamic regulatory mechanisms underlying complex biological processes.
\end{abstract}

\section{Introduction}

Identification of key regulatory genes is fundamental for understanding the dynamic mechanisms underlying complex biological processes, such as development, immune response, and tumor evolution\cite{Gene1,Gene2,Gene3}. Currently, genome-wide association studies and differential expression analyses have identified numerous genes associated with diseases or phenotypes\cite{DEGs1,DEGs2,DEGs3}. However, these methods primarily rely on statistical associations within static snapshots; while they identify participants in biological processes, they struggle to distinguish the drivers of state transitions. Moreover, gene function is not static but rather exhibits time-varying network regulatory effects across different stages. Therefore, moving beyond simple correlation to pinpoint driver genes that govern biological changes within specific temporal windows is necessary for dissecting biological mechanisms. Accurately capturing these drivers not only helps deconstruct cell fate determination mechanisms but also reveals core regulators that direct phenotypic changes and serve as potential therapeutic targets.

Single-cell sequencing technologies have enabled high-resolution characterization of the continuous manifold of cell states\cite{10xGenomics,AvivRegev}. Nevertheless, existing dynamic analysis methods face two major limitations in quantifying driver effects. First, methods represented by RNA Velocity\cite{velocity,scVelo} and optimal transport\cite{OTinSinglecell1,OTinSinglecell2} excel at reconstructing cellular trajectories and geometric flows. However, they typically focus on describing the overall direction and velocity of cell population trajectories, often failing to elucidate the underlying causal gene regulation driving these directional changes. Second, methods for master regulator identification based on gene regulatory networks (GRNs)\cite{TF1,TF2,TF3}, while focusing on gene-gene interactions, rely mostly on static networks or pseudotime approximations. These approaches often aim to identify source factors initiating the entire process, making it difficult to flexibly capture time-varying, stage-specific driver effects dominated by different factors during multi-stage evolution.

To address these challenges, we present Time-varying Network Driver Estimation (TNDE), a computational framework based on optimal transport and graph neural networks. TNDE aims to reconstruct the topology of time-varying GRNs from single-cell snapshots and identify driver effects within arbitrary time windows based on network connectivity strength. The algorithm first utilizes a graph attention autoencoder to map high-dimensional data onto a low-dimensional manifold, capturing topological features. Subsequently, it introduces partial optimal transport with virtual sink/source nodes to reconstruct cross-stage cellular evolutionary trajectories under assumptions allowing for proliferation and apoptosis. Finally, TNDE decouples time-varying regulatory matrices via weighted ridge regression to directly quantify driver effects. Furthermore, based on the Markov process assumption, the framework supports the flexible evaluation of cumulative driver effects across arbitrary spans through matrix multiplication.

We comprehensively evaluated the performance of TNDE through simulated experiments and empirical data analysis. Simulation results demonstrate that TNDE outperforms ablation variants and conventional trajectory alignment baselines in identifying time-varying driver genes. Notably, TNDE exhibits high accuracy and robustness in complex scenarios where driver effects reverse or exist only within specific windows. Furthermore, applied to single-cell transcriptomic data of mouse erythropoiesis, TNDE not only pinpointed stage-specific key regulators but also revealed regulatory mechanisms during differentiation by identifying key genes with significantly downregulated expression yet strong network driving force. These results confirm that TNDE effectively addresses the limitation in capturing stage-specific regulatory effects, providing a powerful computational tool for dissecting the dynamic regulatory mechanisms of complex biological processes.

\begin{figure}[htbp]
    \centering
    \includegraphics[width=0.9\linewidth]{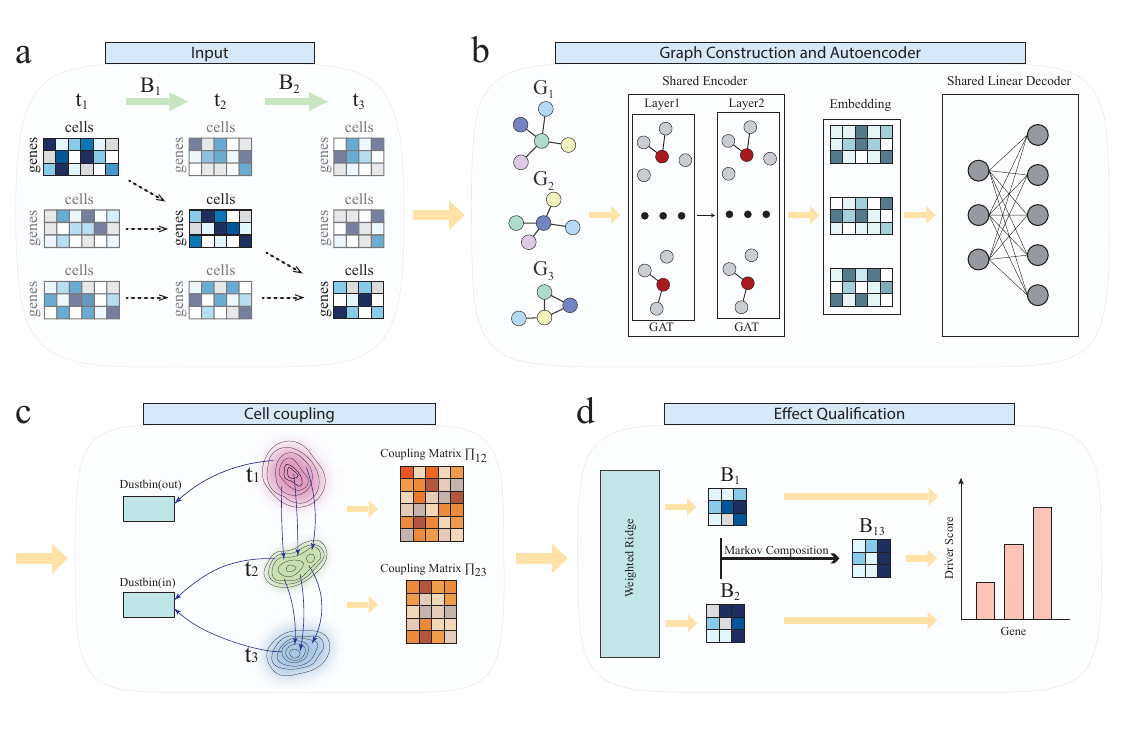}
    \caption{\textbf{Overview of TNDE.}
    \textbf{a}, The framework takes discrete time-series single-cell snapshots as input to uncover the gene regulatory matrices.
    \textbf{b}, A shared Graph Attention Encoder projects high-dimensional gene expression profiles into a common low-dimensional space, preserving the local topological structure of cell neighborhoods.
    \textbf{c}, Partial optimal transport then estimates cell couplings within this space, incorporating a dustbin mechanism to handle unmatched cells from proliferation, apoptosis, or outliers.
    \textbf{d}, Using these couplings, weighted ridge regression infers the gene regulatory network to quantify each gene's driver strength, extending the analysis to long-term cumulative effects via Markov composition.}
    
    \label{fig:overview}
\end{figure}

\section{Results}

\subsection{Overview of the TNDE framework}

We developed the Time-varying Network Driver Estimation (TNDE) framework to recover dynamic gene regulatory networks from discrete single-cell snapshots. Given that single-cell sequencing data provide only static snapshots at discrete time points without continuous lineage tracking, the model posits that observed cells at a later stage evolve from a subset of cells at the preceding stage via a gene regulatory transition matrix. However, accounting for lineage discontinuities caused by cell proliferation, apoptosis, or migration, the cell population at the subsequent stage is modeled as evolving from a combination of observed subpopulations from the current stage and unobserved latent sources (Figure \ref{fig:overview}a).

To address the high dimensionality of single-cell data, TNDE first employs a shared-parameter Graph Attention Network (GAT) to construct low-dimensional representations of the data (Figure \ref{fig:overview}b). This module aggregates local topological information based on the $k$-nearest neighbor graph of cells, effectively suppressing noise while preserving the manifold structure of the cell population, thereby ensuring biologically robust inter-cellular distances in the latent space. Subsequently, to resolve the lineage alignment challenge in non-equilibrium processes, TNDE utilizes partial optimal transport to estimate cross-stage cellular flows within this latent space (Figure \ref{fig:overview}c). Recognizing the discontinuities introduced by proliferation and apoptosis, the algorithm avoids forced global matching; instead, it connects only cells with clear evolutionary relationships, assigning unmatched samples to virtual dustbins. This strategy effectively mitigates the interference of outliers, ensuring the generated cell coupling matrix reflects genuine biological evolutionary paths. Upon obtaining cellular correspondences, TNDE reconstructs the gene regulatory matrix via weighted ridge regression (Figure \ref{fig:overview}d). Within this matrix, the column norm corresponding to each gene quantifies its driver effect on regulatory state transitions. Furthermore, based on the Markov assumption, the framework enables the calculation of cumulative regulatory effects across multiple time points by cascading the transition matrices of consecutive stages.

\subsection{Experimental Design and Datasets}

To validate the accuracy and biological applicability of TNDE in quantifying driver effects, we adopted a comprehensive evaluation strategy integrating both numerical simulations and empirical biological data. Given the lack of observable ground-truth dynamic regulatory factors and weights in real-world single-cell sequencing data, we first constructed a controlled Markov linear Gaussian simulation framework to quantitatively benchmark algorithmic performance. This framework recapitulates challenges in single-cell dynamic analysis, specifically lineage imbalance and the time-varying nature of gene regulatory networks. We incorporated controllable mechanisms for driver gene turnover and effect injection across consecutive stages. This design was aimed at assessing the algorithm's capacity to distinguish time-varying regulatory signals from background co-expression, thereby precisely pinpointing local regulatory alterations occurring within specific temporal windows.

Building on this, to validate the model's generalizability to complex real-world biological scenarios, we applied TNDE to single-cell transcriptomic data of mouse erythropoiesis. This dataset captures the continuous differentiation trajectory from hematopoietic progenitors to committed erythrocytes. As this biological process is characterized by highly stage-specific regulation, we leveraged this dataset to test whether TNDE, in the absence of prior lineage knowledge, could accurately identify driver genes that dominate specific differentiation windows—particularly those often overlooked by static differential expression analyses—by reconstructing the time-varying network topology.

\subsection{TNDE outperforms baselines in identifying dynamic drivers}

To comprehensively evaluate TNDE's performance, we generated simulated data containing three consecutive time points ($t_1, t_2, t_3$) based on a linear Gaussian Markov process. To mimic real biological regulatory complexity, we designed three representative dynamic patterns: the Same mode, where driver genes and their effect strengths remain constant over time; the Partial mode, where the driver gene set evolves but retains a core intersection; and the Disjoint mode, where different stages are driven by completely distinct gene sets. We benchmarked TNDE against six algorithms: ablation variants designed to validate core components (No-Enc, No-OT, kNN), and existing representative trajectory alignment tools (MNN, PCA, Diffusion). Crucially, all methods were tested under identical settings with $p=100$ genes and 10 true driver genes to ensure fair comparison.

\begin{figure}[htbp]
    \centering
    \includegraphics[width=0.9\linewidth]{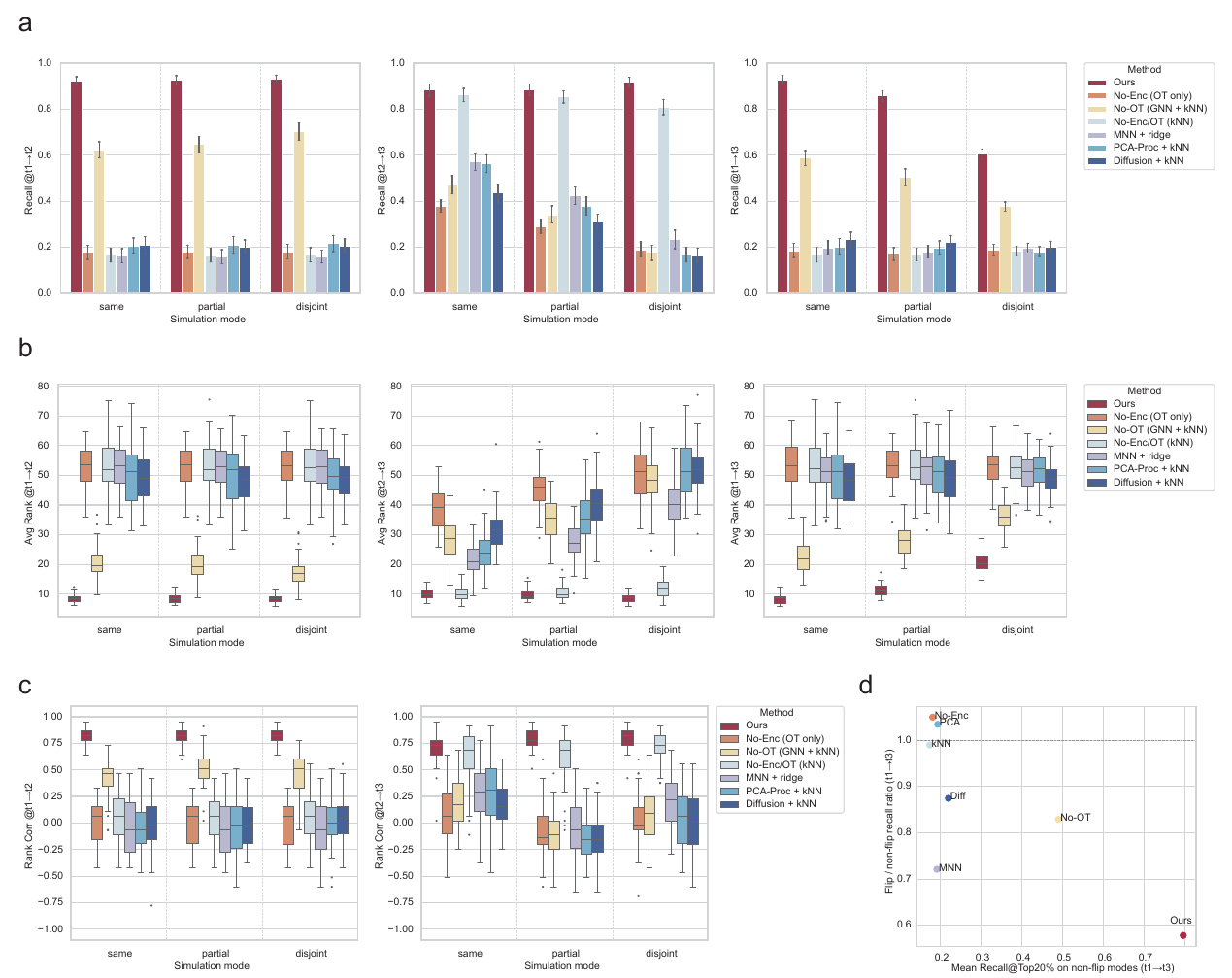}
        \caption{\textbf{Benchmarking TNDE performance on three-stage simulated datasets.}
        \textbf{a}, Bar plots showing the Top-10 recall rates for the first stage, second stage, and the cumulative process under three fundamental dynamic modes.
        \textbf{b}, Box plots displaying the distribution of average ranks for true driver genes across different dynamic modes; lower ranks indicate more precise identification.
        \textbf{c}, Dot plots illustrating the rank correlation coefficients between estimated driver scores and true regulatory strengths in the first and second stages under three dynamic modes.
        \textbf{d}, Scatter plot analyzing the impact of effect reversal on cumulative driver identification. The X-axis represents the average recall in non-reversal modes, while the Y-axis represents the ratio of performance in reversal mode to non-reversal mode; a lower ratio reflects the model's ability to correctly capture the theoretical signal cancellation inherent in cumulative reversal effects.}
    \label{fig:performance}
\end{figure}

TNDE demonstrated consistent superiority across all evaluation metrics (Figure \ref{fig:performance}). Specifically, in terms of Top-N recall, TNDE significantly outperformed other baseline models in both single-stage transitions ($t_1 \to t_2, t_2 \to t_3$) and the cumulative process ($t_1 \to t_3$), exhibiting robust driver gene identification capabilities (Figure \ref{fig:performance}a). Notably, even in the Disjoint mode where the driver landscape changes fundamentally, TNDE maintained accurate identification, whereas static baselines like MNN and PCA showed significant performance degradation. The average rank metric further corroborated this advantage, with TNDE consistently ranking true driver genes higher than any other method (Figure \ref{fig:performance}b). Moreover, rank correlation analysis for single stages revealed a high degree of consistency between TNDE-estimated driver scores and true regulatory strengths, indicating the algorithm's ability to not only pinpoint key genes but also accurately quantify their relative regulatory potency (Figure \ref{fig:performance}c).

This performance advantage underscores the necessity of TNDE's core architectural design. Comparison with ablation variants revealed that removing the GNN encoder (No-Enc) led to a marked performance decline, highlighting the critical importance of capturing non-linear topological structures within cell neighborhoods. Similarly, the variant relying on simple kNN matching (No-OT) performed significantly worse than TNDE, confirming the robustness of partial optimal transport in handling non-equilibrium processes by preventing forced misalignments, thereby ensuring the accuracy of downstream regulatory inference.

To further demonstrate our algorithm's performance in capturing cumulative driver effects across multiple time points, we simulated a specific scenario where the sign of a driver gene's regulation flips between stages. As shown in Figure \ref{fig:performance}d, TNDE resides in the bottom-right quadrant, indicating that it not only maintains high accuracy in standard modes but also sensitively identifies signal cancellation in the reversal mode. In contrast, baseline methods typically failed to distinguish signal cancellation from noise. Collectively, these results demonstrate that TNDE effectively addresses the limitations of existing methods in quantifying driver effects by integrating graph representation learning with optimal transport-based cell alignment.

\subsection{TNDE uncovers stage-specific and latent regulatory drivers in mouse erythropoiesis}

\begin{figure}[htbp]
    \centering
    \includegraphics[width=0.9\linewidth]{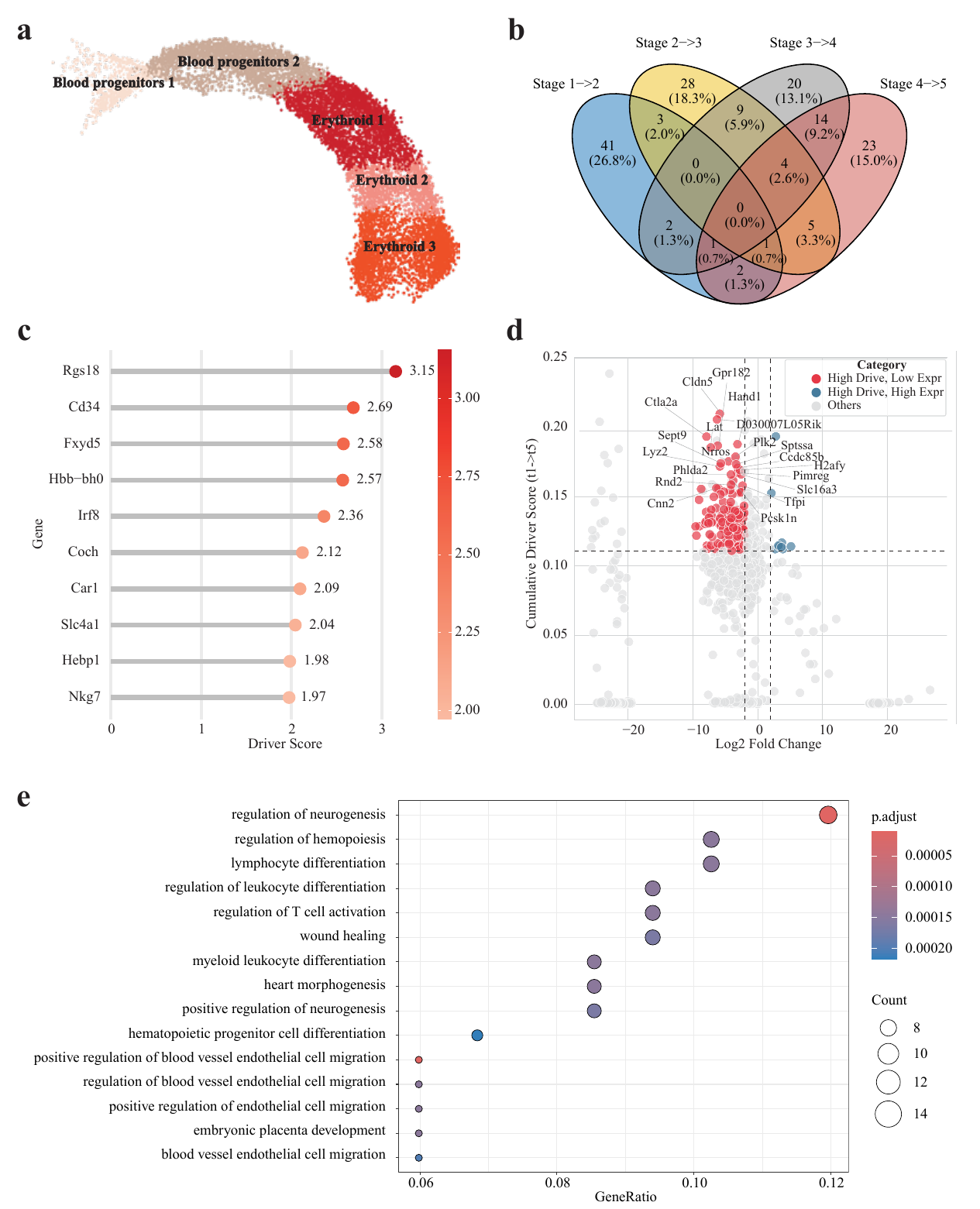}
        \caption{\textbf{Application of TNDE to mouse erythropoiesis data.}
        \textbf{a}, UMAP projection visualizing the cellular developmental trajectory, color-coded to indicate five consecutive differentiation stages from hematopoietic progenitors to erythrocytes.
        \textbf{b}, Venn diagrams quantifying the number of identified driver genes and their overlap ratios across transition stages.
        \textbf{c}, Lollipop chart listing the top 10 key driver genes ranked by cumulative driver score across the entire $t_1 \to t_5$ process.
        \textbf{d}, Scatter plot of cumulative driver scores (Y-axis) versus Log2 fold change in expression (X-axis). Red points mark genes identified by TNDE as high-scoring (Top 25\%) and significantly downregulated (Log2FC $< -1$, $p<0.05$); blue points mark high-scoring (Top 25\%) and significantly upregulated (Log2FC $> 1$, $p<0.05$) genes.
        \textbf{e}, Bubble plot showing the top 15 significant GO enrichment results for high-scoring driver genes. Dot size represents the number of enriched genes, and color intensity indicates statistical significance (Adjusted $p$-value).}
    \label{fig:emperical}
\end{figure}

To further explore TNDE's capacity to dissect complex biological dynamics, we applied it to a single-cell transcriptomic dataset of mouse erythropoiesis \cite{Dataset}. As shown in the Uniform Manifold Approximation and Projection (UMAP) visualization (Figure \ref{fig:emperical}a), this dataset captures a continuous differentiation trajectory from early hematopoietic progenitors to committed erythrocytes. We divided this trajectory into four consecutive transition stages ($t_1 \to t_2$ to $t_4 \to t_5$) to identify dynamic driver genes. TNDE analysis revealed a highly dynamic distribution of driver genes. Venn diagrams of driver genes across the four transition stages showed limited overlap between adjacent stages (Figure \ref{fig:emperical}b). This result suggests that distinct regulatory programs are sequentially activated during lineage commitment and maturation, a temporal specificity that static network inference methods struggle to capture.

Furthermore, the top driver genes identified by TNDE aligned highly with existing biological knowledge. As shown in Figure \ref{fig:emperical}c, the set of top-ranked driver genes clearly delineates the underlying regulatory trajectory of erythroid differentiation. First, the list includes the classic early hematopoietic stem/progenitor cell marker \textit{Cd34} \cite{CD34_1,CD34_2}. More importantly, the top-ranked \textit{Rgs18} is a confirmed myeloid-erythroid lineage-specific regulator \cite{Rgs18_1}, highly enriched in hematopoietic stem cells and early progenitors, and is critical for fine-tuning G-protein signaling pathways to determine lineage fate \cite{Rgs18_2}. As differentiation progresses, the list also includes multiple downstream effector genes closely related to erythrocyte maturation and function, including \textit{Hbb-bh0} (hemoglobin synthesis), \textit{Hebp1} (heme metabolism), and the erythrocyte membrane transporters \textit{Slc4a1} and \textit{Car1}. TNDE's ability to accurately capture this series of key genes, covering the entire process from early commitment to terminal differentiation, strongly validates the biological relevance of the model's driver scoring metric.

To investigate the relationship between driver effects and gene expression changes, we compared the cumulative driver scores of all genes against their Log2 fold changes in expression (Figure \ref{fig:emperical}d). The analysis revealed that high-weight drivers identified by TNDE predominantly exhibited significant downregulation in expression (Log2FC < -1, p < 0.05). This finding reveals a unique regulatory mechanism at this developmental stage: erythroid lineage commitment is driven largely by the repression of key genes maintaining the early progenitor state, rather than solely by the activation of new genes. Furthermore, we performed Gene Ontology (GO) enrichment analysis on this set of genes with high driver scores but significant downregulation to understand their functional characteristics (Figure \ref{fig:emperical}e). Results showed that these genes were significantly enriched in terms such as "regulation of hemopoiesis" and "hematopoietic progenitor cell differentiation." This confirms that TNDE accurately captured the critical signals where cells determine lineage fate by downregulating specific pluripotency regulators upon exiting the progenitor state. Additionally, enrichment terms such as "regulation of blood vessel endothelial cell migration" further corroborates the complex spatiotemporal regulatory roles of this gene set during early cellular development.

\section{Methods}
\subsection{Data Simulation}
\label{sec:simulation}
To systematically evaluate the capacity of TNDE to identify driver effects, we designed a controllable simulator based on a Markov linear Gaussian process. This simulator recapitulates two key features of real biological data: (i) time-varying regulatory effects that evolve dynamically; and (ii) partially coupled populations arising from mixed cellular lineages.

\vspace{4pt}
\noindent\textbf{Generation of Regulatory Matrices}

We first generated ground-truth cross-stage regulatory matrices $B_1$ ($t_1 \to t_2$) and $B_2$ ($t_2 \to t_3$). We modeled the gene regulatory network as a superposition of a shared basal network and a stage-specific network. Thus, each $B_t$ was formulated as a combination of a common regulatory component $C$ and a specific regulatory component $\Delta_t$:
\begin{equation}
B_t = (1-w) \Delta_t + w C,\qquad w \in [0, 1]\
\end{equation}

\vspace{4pt}
\noindent\textbf{Injection of Driver Effects}

Subsequently, we injected predefined driver effects into the regulatory matrix $B_t$. In our model, $B_{:,k}$ represents the regulatory strength of gene $k$ on all other genes. We set the $L_2$ norm of the column $B_{:,k}$ corresponding to the selected driver gene $k$ to a predefined driver strength. To test the robustness of the algorithm under different dynamic scenarios, we established four driver gene modes defining the relationship between the driver gene sets in $B_1$ and $B_2$:
\begin{itemize}
    \item \textbf{Same:} The set of driver genes and their effect strengths remain identical across both stages. 
    \item \textbf{Disjoint:} The sets of driver genes in the two stages are completely mutually exclusive. 
    \item \textbf{Partial:} The sets of driver genes evolve over time but retain a partial intersection. 
    \item \textbf{Flip:} The set of driver genes remains the same, but their regulatory effects are completely reversed (sign-flipped) in the second stage. 
\end{itemize}

\vspace{4pt}
\noindent\textbf{Multi-stage Data Generation}

We assumed a basal cell population $X_1 \sim \mathcal{N}(0, I)$ at $t=1$. Observational data $X_t$ at subsequent stages were modeled as a mixture of two distinct source populations to simulate cell groups with different lineage origins but partial associations.

Let $m_t$ be the linkage proportion from $t-1 \to t$. The dataset $X_t$ comprises $n_t m_t$ cells constituting the linked population ($X_{t, \text{linked}}$). The generation of this population follows a Markov process evolving from $X_{t-1}$:
\begin{equation}
X_{t, \text{linked}} = X_{t-1}[\mathcal{P}_t]\, B_{t-1}^\top + \varepsilon_1,
\end{equation}
where $\mathcal{P}_t \subset \{1, \dots, n_{t-1}\}$ is a set of parent cell indices with size $n_t m_t$, and $\varepsilon_1 \sim \mathcal{N}(0, \sigma^2 I)$ represents Gaussian noise.

The remaining $n_t(1-m_t)$ cells constitute the unlinked population ($X_{t, \text{unlinked}}$). These cells follow the same Markov transition defined by $B_{t-1}$, but their evolutionary origin differs. They evolve from a newly generated basal distribution $X_{t, \text{root}} \sim \mathcal{N}(0, I)$, independent of $X_{t-1}$:
\begin{equation}
X_{t, \text{unlinked}} = X_{t, \text{root}}\, B_{t-1}^\top + \varepsilon_2.
\end{equation}

The finally observed $X_t$ is a mixture of $X_{t, \text{linked}}$ and $X_{t, \text{unlinked}}$. By recursively applying this process, we generated a dataset containing multiple time points $\{X_1, X_2, \dots, X_t\}$.

\subsection{The TNDE Algorithm}

\vspace{4pt}
\noindent\textbf{Data Preprocessing}

Let the data at stage $t$ be $X_t \in \mathbb{R}^{n_t \times p}$, where $n_t$ is the number of cells and $p$ is the number of genes. As a preprocessing step, we calculated the global mean and variance across all stages $\{X_1, \dots, X_T\}$ to perform global standardization. All subsequent analyses were conducted on these standardized data.

\subsubsection{Shared Graph Attention Encoder}
\vspace{4pt}
\noindent\textbf{Graph Topology Construction}

To characterize inter-cellular relationships, we first constructed a $k$-nearest neighbor undirected graph $\mathcal{G}_t=(\mathcal{V}_t,\mathcal{E}_t)$ based on Euclidean distances between samples, yielding an adjacency matrix $A_t\in\{0,1\}^{n_t\times n_t}$. To facilitate graph signal processing, we computed the symmetric normalized adjacency matrix $\tilde{A}_t=D_t^{-1/2}A_tD_t^{-1/2}$ and the graph Laplacian $L_t=I-\tilde{A}_t$.

\vspace{4pt}
\noindent\textbf{Shared Encoder Architecture}

Leveraging this graph structure, we designed a parameter-shared Graph Attention Network (GAT) encoder to learn low-dimensional representations $Z_t\in\mathbb{R}^{n_t\times d_z}$ that preserve the topological relationships among cells. The encoder consists of $L$ stacked Transformer blocks followed by a linear projection layer. Each Transformer block contains two sub-layers: (i) a graph-masked multi-head attention layer and (ii) a point-wise feed-forward network (FFN). Each sub-layer is accompanied by residual connections and layer normalization. Denoting the hidden dimension as $d$, the forward propagation for each stage $t$ is defined as:
\begin{align}
H_t^{(0)}&=\phi\!\big(X_t W_{\text{in}}\big),\\
H_t^{(\ell+\tfrac{1}{2})}&=\mathrm{LN}\!\Big(H_t^{(\ell)}+\mathrm{Attn}\big(H_t^{(\ell)},A_t\big)\Big),\\
H_t^{(\ell+1)}&=\mathrm{LN}\!\Big(H_t^{(\ell+\tfrac{1}{2})}+\mathrm{FFN}\big(H_t^{(\ell+\tfrac{1}{2})}\big)\Big),\\
Z_t&=g\!\big(H_t^{(L)}\big),
\end{align}
where $\phi$ is the GELU activation function, $\mathrm{LN}$ denotes layer normalization, $\mathrm{FFN}$ is a two-layer point-wise feed-forward network, and $g(\cdot)$ is a linear map to dimension $d_z$.

In the attention sub-layer, for input $H\in\mathbb{R}^{n\times d}$, the $h$-th attention head is computed as:
\begin{align}
Q^{(h)}&=H W_Q^{(h)},\quad
K^{(h)}=H W_K^{(h)},\quad
V^{(h)}=H W_V^{(h)},\\
s_{ij}^{(h)}&=\frac{\langle q_i^{(h)},k_j^{(h)}\rangle}{\sqrt{d_h}},\qquad
\alpha_{ij}^{(h)}=\frac{\exp\big(s_{ij}^{(h)}\big)\cdot \mathbb{I}\{A_{ij}=1\}}{\sum_{j':A_{ij'}=1}\exp\big(s_{ij'}^{(h)}\big)},
\end{align}
The outputs of all heads $U^{(h)}$ are concatenated and linearly projected by $W_O$ to obtain the sub-layer output:
\begin{equation}
\mathrm{Attn}(H,A)=\Big[U^{(1)}\ \Vert\ \cdots\ \Vert\ U^{(H)}\Big]\,W_O,\qquad W_O\in\mathbb{R}^{(Hd_h)\times d}.
\end{equation}

\vspace{4pt}
\noindent\textbf{Loss Function}

To ensure the low-dimensional embeddings can be mapped back to the gene space for interpretation, we introduced a shared linear decoder $W_{\text{dec}}\in\mathbb{R}^{d_z\times p}$ for each stage:
\begin{align}
\hat{X}_t &= Z_t W_{\text{dec}},\qquad
\mathcal{L}_{\text{recon}}(X_t,Z_t)=\frac{1}{n_t}\,\|X_t-\hat{X}_t\|_F^2.
\end{align}
Simultaneously, to suppress local noise and preserve neighborhood geometry, we applied Laplacian smoothing regularization to each stage. We incorporated warm-up and adaptive normalization to balance the magnitude differences between penalty terms:
\begin{equation}
\mathcal{L}_{\text{smooth}}(Z_t)=\frac{1}{n_t}\,\mathrm{tr}\!\big(Z_t^\top L_t Z_t\big)
=\frac{1}{2n_t}\sum_{i,j}\tilde{A}_{t,ij}\,\|z_i-z_j\|_2^2.
\end{equation}
\begin{equation}
\mathcal{L}_{\text{smooth}}^{\text{(eff)}}(k)
=\gamma_{\text{smooth}}\cdot
\underbrace{\min\!\Big(1,\tfrac{k}{K_w}\Big)}_{\text{warm-up}}
\cdot
\underbrace{\frac{\mathcal{L}_{\text{smooth}}^{(k)}}{\mathcal{L}_{\text{smooth}}^{(0)}+\epsilon}}_{\text{adaptive norm}},
\end{equation}
where $k$ is the training epoch, $K_w$ is the number of warm-up epochs, and $\epsilon$ is a constant for numerical stability.

Thus, the final optimization objective is:
\begin{equation}
\min_{\Theta}
\ \sum_{t=1}^{T}\Big[
\underbrace{\mathcal{L}_{\text{recon}}(X_t,Z_t)}_{\text{Reconstruction Loss}}
\ +\ 
\underbrace{\mathcal{L}_{\text{smooth}}^{\text{(eff)}}(k)}_{\text{Graph Smoothing Loss}}
\Big],
\end{equation}
where $\Theta$ denotes the shared parameters of the encoder and decoder.

\subsubsection{Optimal Transport with Dustbin Extension}
To establish correspondences between embeddings $Z_s$ and $Z_t$ of adjacent stages, while accounting for unmatched scenarios such as cell birth, apoptosis, or outliers, we adopted a partial optimal transport mechanism. This allows allocating a fraction of probability mass to virtual nodes, thereby avoiding forced global matching.

\vspace{4pt}
\noindent\textbf{Joint Whitening}

Before computing matching, to mitigate metric mismatch caused by covariance shift, we applied a joint whitening strategy. We first concatenated the embeddings of two stages along the sample axis to compute a joint mean $\mu_{\text{joint}}$ and covariance $C_{\text{joint}}$:
\begin{equation}
\mu_{\text{joint}}=\tfrac{1}{n_s+n_t}\mathbf{1}^\top [Z_s;Z_t],\quad
C_{\text{joint}}=\tfrac{1}{n_s+n_t-1}\big([Z_s;Z_t]-\mu_{\text{joint}}\big)^\top\big([Z_s;Z_t]-\mu_{\text{joint}}\big).
\end{equation}
Subsequently, we centered and transformed $Z_s$ and $Z_t$ using the same whitening matrix $W=(C_{\text{joint}}+\rho I)^{-1/2}$ ($\rho>0$):
\begin{equation}
Z_s^{\mathrm{w}}=(Z_s-\mu_{\text{joint}})\,W,\qquad
Z_t^{\mathrm{w}}=(Z_t-\mu_{\text{joint}})\,W.
\end{equation}

\vspace{4pt}
\noindent\textbf{Cost Matrix}

In the whitened space, we constructed the transport cost matrix using squared Euclidean distances:
\begin{equation}
C_{ij}=\big\| z_i^{(s),\mathrm{w}}-z_j^{(t),\mathrm{w}}\big\|_2^2,\qquad 1\le i\le n_s,\ 1\le j\le n_t .
\end{equation}
To enable partial matching, we extended $C$ with virtual dustbins, yielding an augmented cost matrix:
\begin{equation}
C^{\mathrm{aug}}=
\begin{bmatrix}
C & \delta_{\mathrm{out}}\mathbf{1}_{n_s}\\[2pt]
\delta_{\mathrm{in}}\mathbf{1}_{n_t}^\top & 0
\end{bmatrix}\in\mathbb{R}^{(n_s+1)\times(n_t+1)},
\end{equation}
where $\delta_{\mathrm{out}}$ and $\delta_{\mathrm{in}}$ represent the penalty intensities for entering or leaving the set of real samples, respectively.

Let $m\in(0,1]$ be the fraction of total mass to be matched. The corresponding augmented marginal distributions are defined as:
\begin{equation}
a^{\mathrm{aug}}=\big(\tfrac{m}{n_s}\mathbf{1}_{n_s};\ 1-m\big),\qquad
b^{\mathrm{aug}}=\big(\tfrac{m}{n_t}\mathbf{1}_{n_t};\ 1-m\big).
\end{equation}

\vspace{4pt}
\noindent\textbf{Entropy-Regularized Solution}

We formulated this as an entropy-regularized partial optimal transport problem:
\begin{equation}
\min_{P\in\mathbb{R}_+^{(n_s+1)\times(n_t+1)}}
\ \ \langle P, C^{\mathrm{aug}}\rangle - \varepsilon\, H(P)
\quad ,\varepsilon>0\ \quad \text{s.t.}\quad
P\mathbf{1}=a^{\mathrm{aug}},\ \ P^\top\mathbf{1}=b^{\mathrm{aug}},
\end{equation}
where $H(P)=-\sum_{ij}P_{ij}\log P_{ij}$ is the Shannon entropy.

This convex optimization problem can be solved by iteratively scaling the Gibbs kernel $K=\exp(-C^{\mathrm{aug}}/\varepsilon)$ to satisfy marginal constraints $a^{\mathrm{aug}}$ and $b^{\mathrm{aug}}$ (Sinkhorn algorithm). The optimal solution $P^\star$ takes the form:
\begin{equation}
P^\star=\mathrm{diag}(u)\,K\,\mathrm{diag}(v),
\end{equation}
where $u,v$ are the scaling vectors obtained upon convergence.

Finally, we extracted the top-left sample-to-sample sub-block of $P^\star$ as the coupling matrix between stages:
\begin{equation}
\Pi = P^\star_{1:n_s,\ 1:n_t}\in\mathbb{R}_+^{n_s\times n_t}.
\end{equation}

\subsubsection{Regulatory Estimation and Multi-step Driver Identification}
\label{sec:ridge-markov}

\vspace{4pt}
\noindent\textbf{Estimation of Regulatory Relationships}

Given observations from adjacent stages $(X_s, X_t)$ and their coupling matrix $\Pi$, we estimated the gene regulatory matrix $B\in\mathbb{R}^{p\times p}$ from $X_s$ to $X_t$ using weighted least squares with ridge regularization:
\begin{equation}
\min_{B\in\mathbb{R}^{p\times p}}\ \
\sum_{i=1}^{n_s}\sum_{j=1}^{n_t}\Pi_{ij}\,\big\|x_j^{(t)}-x_i^{(s)}B^\top\big\|_2^2
\;+\; \lambda \,\|B\|_F^2,\qquad \lambda>0.
\end{equation}
Letting $r=\Pi\mathbf{1}\in\mathbb{R}^{n_s}$ and $D_r=\mathrm{diag}(r)$, and defining:
\begin{equation}
A = X_s^\top D_r X_s\in\mathbb{R}^{p\times p},\qquad
S = X_s^\top \Pi X_t\in\mathbb{R}^{p\times p},
\end{equation}
the solution is given by:
\begin{equation}
\big(A+\lambda I\big)\,B^\top = S
\quad\Rightarrow\quad
B^\top = \big(A+\lambda I\big)^{-1} S .
\end{equation}

\vspace{4pt}
\noindent\textbf{Identification of Driver Genes}

For any gene regulatory matrix, $B_{:,k}$ represents the aggregate regulatory effect of gene $k$ on all other genes during this transition. Therefore, we defined the driver effect score of gene $k$ as the $L_2$ norm of its corresponding column:
\begin{equation}
\text{Score}(k) = \|\hat{B}_{:, k}\|_2.
\end{equation}
A higher score indicates a stronger regulatory driver effect of gene $k$ during this stage transition.

Furthermore, assuming the cellular transition process follows a first-order linear Gaussian Markov model:
\begin{equation}
X_{t+1}=X_t B_t^\top+\varepsilon_t,\qquad \mathbb{E}[\varepsilon_t]=0,\ \ \varepsilon_t\ \perp\ X_t,
\end{equation}
it follows that
\begin{equation}
\mathbb{E}[X_{t_2}\mid X_{t_1}]=X_{t_1}\big(B_{t_2-1}\cdots B_{t_1}\big)^\top.
\end{equation}
Based on this Markov assumption, the estimated matrices from adjacent stages $\{B_{t}\}$ can be combined to form a global regulatory matrix for any continuous interval:
\begin{equation}
B_{t_1\!\to t_q}\ =\ B_{t_q-1}\,B_{t_q-2}\cdots B_{t_1}\ \in\ \mathbb{R}^{p\times p}.
\end{equation}
This matrix characterizes the composite linear effect from $t_1$ to $t_q$.
This formulation enables us to identify not only stage-wise driver genes by analyzing the column norms of $\hat{B}_t$, but also cumulative driver effects over arbitrary time scales by computing the column norms of the composite matrix.

\section{Discussion}

While single-cell RNA sequencing (scRNA-seq) provides high-resolution snapshots of cell states, dissecting dynamic regulatory mechanisms from these static images remains challenging. In this study, we present the TNDE framework, a computational model integrating graph representation learning, partial optimal transport, and sparse regression. Distinct from existing trajectory inference or static network analysis methods, the innovation of TNDE lies in explicitly modeling cell state transitions as a non-equilibrium transport process accompanied by proliferation and apoptosis, thereby quantifying time-varying gene driver effects on this basis.

The superior performance of TNDE is attributed to its precise modeling of the intrinsic characteristics of single-cell data. First, simulation experiments demonstrate that traditional alignment methods based on kNN or standard optimal transport  are prone to forced erroneous matching when dealing with lineage branching or discontinuous sampling. In contrast, TNDE introduces a virtual node mechanism to automatically identify and exclude unexplainable cells. This alignment strategy significantly enhances the robustness of subsequent gene regulatory network inference, particularly in non-stationary processes where cellular population structures undergo drastic changes. Second, by utilizing a shared graph attention encoder, we effectively suppress sequencing noise while preserving local topological structures, rendering regulatory inference on the low-dimensional manifold more reliable.

In terms of biological application, analysis of mouse erythropoiesis data revealed that the majority of high-scoring driver genes identified by TNDE were significantly downregulated in differential expression analysis. This indicates that TNDE quantifies driver effects directly through the reconstructed gene regulatory network, rather than relying solely on the magnitude or direction of expression changes. This network topology-based quantification enabled TNDE to accurately identify key progenitor factors that drive cell fate transitions through downregulation, validating the model's effectiveness in capturing complex regulatory patterns.

Despite TNDE's robust performance in capturing dynamic driver genes, this study has limitations inherent to the model assumptions. While weighted ridge regression was adopted to ensure computational stability and interpretability given the sparse and noisy nature of single-cell data, true gene regulatory networks are inherently nonlinear, often involving enzymatic saturation effects, transcriptional activation thresholds, or complex combinatorial logic. Consequently, the current linear model represents a first-order approximation of complex biological dynamics. While this effectively captures major regulatory trends, it may fail to fully resolve fine-grained nonlinear regulatory details. Furthermore, the current framework relies solely on transcriptomic data, limiting the direct inference of cis-regulatory elements at the epigenetic level.

Future work will primarily focus on overcoming these linear constraints. On one hand, we plan to introduce nonlinear dynamic models to replace the current linear regression, thereby fitting the nonlinear evolutionary behavior of complex systems with higher fidelity. On the other hand, we aim to extend TNDE to support the integration of multi-omics data, utilizing chromatin accessibility as a prior structural constraint to further enhance the accuracy and biological resolution of GRN inference. In summary, TNDE provides a powerful computational tool for dissecting dynamic regulatory mechanisms in development and disease progression, laying a methodological foundation for future precision medicine research.

\section*{Code and Data Availability}
The complete source code and associated datasets supporting this study are publicly accessible at \url{https://github.com/jxlee423/TNDE}

\end{document}